\documentclass[10pt,letterpaper]{article}
\usepackage{opex3}
\usepackage{bm}
\usepackage{amsmath}
\usepackage{amstext}
\usepackage{amssymb}
\usepackage{graphicx}

\newcommand{\SSS}{\scriptscriptstyle}

\newcommand{\Ee}{{\rm e}}
\newcommand{\Dd}{{\rm d}}

\newcommand{\Ii}{{\rm i}}

\newcommand{\EpsA}{\epsilon_{1}}
\newcommand{\EpsB}{\epsilon_{2}}
\newcommand{\MuA}{\mu_{1}}
\newcommand{\MuB}{\mu_{2}}

\newcommand{\EtaA}{\eta_{1}}
\newcommand{\EtaB}{\eta_{2}}

\newcommand{\GmA}{\gamma_{1}}
\newcommand{\GmB}{\gamma_{2}}

\newcommand{\EffEpsRd}{\tilde{\epsilon}_{r}}
\newcommand{\EffEpsPh}{\tilde{\epsilon}_{\phi}}

\newcommand{\PhTM}{\Phi_{\SSS\mathrm{E}}}
\newcommand{\PhTE}{\Phi_{\SSS\mathrm{H}}}

\begin{document}

%% START HERE
%%%%%%%%%%%%%%%%%% title page information %%%%%%%%%%%%%%%%%%
\title{Plasmonic angular momentum on metal-dielectric nano-wedges in a sectorial indefinite metamaterial}

\author{Dafei Jin and Nicholas X. Fang$^*$}

\address{Department of Mechanical Engineering, Massachusetts Institute of Technology, \\ Cambridge, MA, 02139, USA}

\email{$^*$nicfang@mit.edu} %% email address is required

% \homepage{http:...} %% author's URL, if desired

%%%%%%%%%%%%%%%%%%% abstract and OCIS codes %%%%%%%%%%%%%%%%
%% [use \begin{abstract*}...\end{abstract*} if exempt from copyright]

\begin{abstract} We present an analytical study to the structure-modulated plasmonic angular momentum trapped on  periodic metal-dielectric nano-wedges in the core region of a sectorial indefinite metamaterial. Employing a transfer-matrix calculation and a conformal-mapping technique, our theory is capable of dealing with realistic configurations of arbitrary sector numbers and rounded wedge tips. We demonstrate that in the deep-subwavelength regime strong electric field carrying high azimuthal variation can exist within only ten-nanometer length scale close to the structural center, and is naturally bounded by a characteristic radius of the order of hundred-nanometer away from the center. These extreme confining properties suggest that the structure under investigation may be superior to the conventional metal-dielectric waveguides or cavities in terms of nanoscale photonic manipulation.
\end{abstract}

 % REPLACE WITH CORRECT OCIS CODES FOR YOUR ARTICLE

%%%%%%%%%%%%%%%%%%%%%%% References %%%%%%%%%%%%%%%%%%%%%%%%%

%%%%%%%%%%%%%%%%%%%%%%%%%%  body  %%%%%%%%%%%%%%%%%%%%%%%%%%
\section{Introduction}

Along with the extensive studies on various metamaterials in the recent years, the so-called
indefinite metamaterials (or hyperbolic metamaterials) have attracted particular attention
\cite{SmithPRL2003,SmolyaninovPRL2010,YaoPNAS2011,YangNaturePhoton2012}. These artificial materials are commonly constructed with multiple metal-dielectric layers (flat or curved, connected or trenched) so that the effective permittivity tensor brings on different signs in different directions, which results in plasmon-polariton-assisted singular density of states \cite{SmolyaninovPRL2010,KrishnamoorthyScience2012}. Such a strange character has been harnessed to achieve hyperlensing that can transmit near-field photonic information to far field \cite{FangScience2005,JacobOE2006,LiuScience2007,LiNatureMater2009}, and to tune the lifetime of quantum emitters placed inside or in contact with these metamaterials \cite{KrishnamoorthyScience2012}.

\begin{figure}[htbp]
\centerline{\includegraphics[scale=0.75]{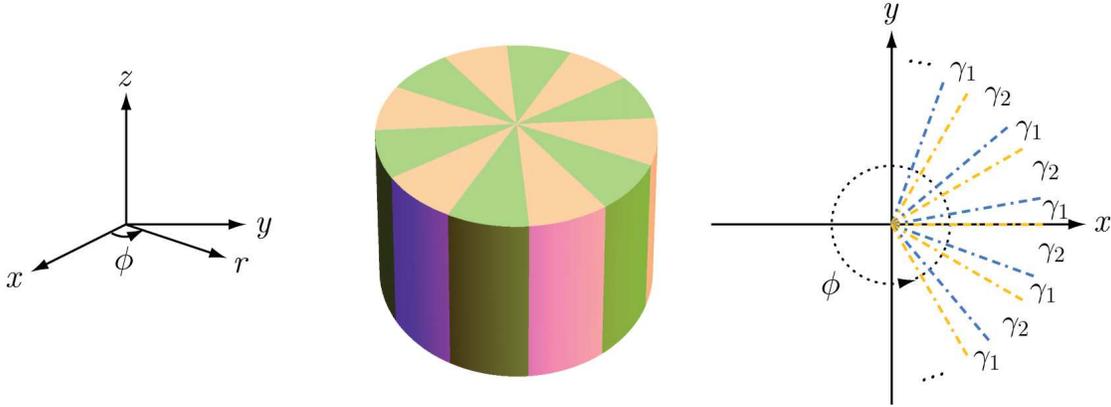}} \caption{Schematics of a sectorial
construction of indefinite metamaterial consisting of two elementary media 1 and 2, periodically
arranged in the azimuthal $\phi$-direction with alternating angular span $\GmA$ and $\GmB$,
respectively.}\label{FigSectorDiagram}
\end{figure}

In this paper, we consider a seemingly basic but far less-studied sectorial construction of indefinite metamaterials, as shown in
Fig.~\ref{FigSectorDiagram}. It consists of two elementary media: 1 (metal) and 2 (
dielectric), periodically arranged in the azimuthal $\phi$-direction, and uniformly extended in the radial $r$-direction and axial $z$-direction in cylindrical coordinates \cite{LiNatureMater2009,LiOE2009}. The angular span of each sector is $\GmA$ or $\GmB$ for the filling medium 1 or 2, respectively. The angular periodicity of one primitive unit (composed of an adjacent pair of medium 1 and medium 2) is $ \gamma=\GmA + \GmB$, and the total unit number
is $N=2\pi/\gamma$. This structure was once proposed by Jacob \textit{et al.} \cite{JacobOE2006} as an alternative hyperlensing construction in parallel with the concentric multilayer one. In their paper, they mainly discuss the hyperlensing functionality in the presence of extrinsic sources, using a two-dimensional effective medium theory (taking the continuous limit $N\rightarrow\infty$, $\gamma\rightarrow 0$, and letting the axial wavenumber $k_z\rightarrow0$). In our work, we focus on the intrinsic so-called plasmonic edge modes \cite{DobrzynskiPRB1972,BoardmanPRB1981,GarciaMolinaPRB1985,MorenoPRL2008} in this structure, under more realistic circumstances when the effective medium theory likely fails. Numerical simulation for similar structures suffering sharp wedge tips often involves instability or inefficiency. By contrast, our transfer-matrix calculation and conformal-mapping technique allow analytically treating arbitrary sector numbers and rounded wedge tips, and can therefore serve as a powerful and reliable toolbox for comprehensive exploration. We are able to systematically compute the eigen-spectrum and field profiles for various structure-modulated \cite{FerrandoPRE2005} plamsonic angular momentum in the deep-subwavelength regime.

Angular momentum of photons has been a topic of great interest for some years, and has found its applications in many fields \cite{AllenPRB2004,PatersonScience2001,MarrucciPRL2006}. It is regarded as a promising candidate for encoding and delivering information in the next-generation optical communication. Angular momentum of plasmons when metals are incorporated into material design has also triggered a lot of interest \cite{KimNanoLett2010,ShenOL2012}. Most studies so far are for relatively long length scale of the order of several hundred nanometers. Making use of the signature of indefinite metamaterials, we demonstrate that in the deep-subwavelength regime the electric field carrying high azimuthal variation can be extremely intense around the structural center where all the nano-sized wedge tips meet. Structure-modulated plasmonic angular momentum in ten-nanometer length scale can form there. For a fixed frequency $\omega$ and a fixed axial wavenumber $k_z$, higher-angular-momentum modes tend to oscillate more drastically and distribute more widely in the radial direction from the structural center. Nevertheless, there always exists a characteristic bounding radius that naturally encapsulates all the field intensity into a region of the order of hundred-nanometer. In comparison, metal-dielectric circular waveguides or cavities following conventional designs are incompetent at confining this high photonic or plasmonic angular momentum in so small length scale, owing to both the geometric and physical restrictions \cite{YehWaveguide2008,HuEPL2012,LiOE2013}. Hence the remarkable properties of the structure under exploration can be potentially useful to the manipulation of photons and plasmons in extreme nanoscale \cite{ChangPRL2006,KlimovPRA2004}.

In the subsequent sections, we will first set up our problem and theoretical framework in a general manner, then present our detailed results and analysis. In the end, we will give a brief summary.

\section{General formalism}

Assume the entire structure to be unbounded in both $r$- and $z$-directions, bearing continuous
translational symmetry along the $z$-axis and discrete rotational symmetry in the $r\phi$-plane.
The permittivities and permeabilities of the two media are $\EpsA, \MuA$ and $\EpsB, \MuB$,
respectively, all of which may depend on the frequency $\omega$. In each sector, the electric field
$\bm{E}$ and the magnetic field $\bm{H}$ take linear combinations of cylindrical waves with the
phase factors like $\Ee^{\Ii\nu\phi} \Ee^{\Ii k_z z}\Ee^{-\Ii\omega t}$, where $k_z$ is the axial
wavenumber in the $z$-direction, $\nu$ is the azimuthal wavenumber in the $\phi$-direction. Owing
to the continuous translational symmetry, $k_z$ is shared by all sectors for any eigenmode of the
whole system, and is real-valued in the absence of any sources that may break the translational
symmetry. A radial wavenumber $k_r$ is related to $\omega$ and $k_z$ by $k_r^2=\mu\epsilon
\omega^2/ c^2 - k_z^2$, where $\epsilon=\EpsA$ or $\EpsB$, $\mu=\MuA$ or $\MuB$, corresponding to
medium 1 or 2. In case the medium is metal ($\mu>0$, $\epsilon<0$), or is dielectric ($\mu>0$,
$\epsilon>0$) but the axial propagation is subwavelength \cite{EconomouPR1969}, we will have
$k_z^2>\mu\epsilon\omega^2/c^2$ and may write $k_r=\Ii\kappa_r$, where $\kappa_r$ is the evanescent radial wavenumber in the $r$-direction,
\begin{equation}
\kappa_r^2= -k_r^2= k_z^2 - \mu\epsilon\frac{\omega^2}{c^2}. \label{RadialWavenumber}
\end{equation}
Moreover, since the continuous rotational symmetry is broken in this structure and each sector is bounded by
two wedge interfaces, $\nu$ can generally take any fractional or even complex numbers. If
$\nu=\Ii\varsigma$ while $\varsigma$ is real-valued, the factors like $\Ee^{\Ii\nu\phi} =
\Ee^{-\varsigma\phi}$ represent azimuthal evanescent waves in the vicinity of wedge interfaces.

We shall treat the eigenmode problem of our structure as a waveguide problem \cite{YehWaveguide2008,JacksonElectrodynamics1998}, in the sense that the
electromagnetic waves of interest are primarily propagating in the longitudinal direction along the
$z$-axis but bounded in the transverse direction in the $r\phi$-plane. (We do not consider problems
of radially outgoing or incoming waves emitted from or scattered by this structure.) Employing a
more convenient representation, we may fully describe the problem with two scalar potentials
$\PhTM$ and $\PhTE$ instead of the more familiar $\bm{E}$ and $\bm{H}$ fields. $\PhTM$ stands for
the $E_z$-waves (or transverse-magnetic waves), and $\PhTE$ stands for the $H_z$-waves (or
transverse-electric waves). They both satisfy the two-dimensional scalar Helmholtz equation,
\begin{align}
\left[\partial_r^2+\frac{1}{r^2}\partial_\phi^2 + \frac{\omega^2}{c^2} \mu\epsilon - k_z^2 \right]
\PhTM = 0 , \qquad \left[\partial_r^2+\frac{1}{r^2}\partial_\phi^2 + \frac{\omega^2}{c^2}
\mu\epsilon - k_z^2\right] \PhTE = 0 .
\end{align}
The general eigenmodes in sectorial structures are necessarily $E_z$-$H_z$-hybridized modes. So the $\PhTM$-$\PhTE$-combined electric and magnetic fields can be generated via
\begin{align}
\bm{E} & = -\nabla \PhTM + \Ii\mu\epsilon \frac{\omega^2}{c^2 k_z} \PhTM
\mathbf{e}_z -\mu\frac{\omega}{c k_z}\nabla \times \left[ \PhTE
\mathbf{e}_z \right],\label{ElectricField} \\
\bm{H} & = -\nabla \PhTE + \Ii\mu\epsilon \frac{\omega^2}{c^2
k_z} \PhTE \mathbf{e}_z +\epsilon\frac{\omega}{c k_z} \nabla \times \left[ \PhTM \mathbf{e}_z
\right],\label{MagneticField}
\end{align}
which contain both the longitudinal and transverse components with respect to the directional unit vector $\mathbf{e}_z$. The boundary conditions across the wedge interfaces
are the continuities of $E_z$, $E_r$, $\epsilon E_\phi$ and $H_z$, $H_r$, $\mu H_\phi$.

\begin{figure}[htbp]
\centerline{\includegraphics[scale=0.75]{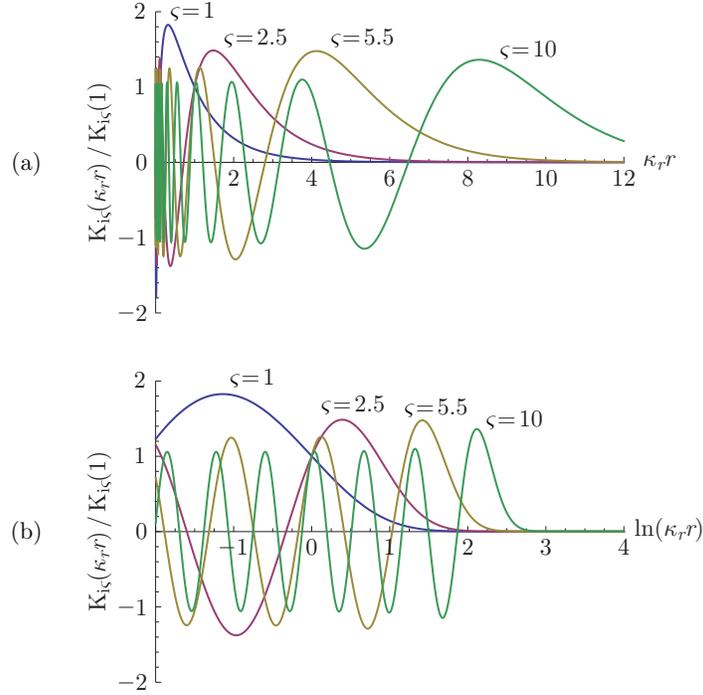}} \caption{Plots of the complex-order
modified Bessel function of the second kind $\mathrm{K}_{\Ii\varsigma}(\kappa_r
r)/\mathrm{K}_{\Ii\varsigma}(1)$ with $\varsigma=1, 2.5, 5.5, 10$, when the abscissa is taken as (a) $\kappa_rr$
and (b) $\ln(\kappa_rr)$. The denominator $\mathrm{K}_{\Ii\varsigma}(1)$ is introduced to cancel some
large prefactors and optimize the visualization.}\label{FigKelvinFunction}
\end{figure}

To solve the waveguide modes in our metal-dielectric construction, $\kappa_r$ must be real-valued
(neglecting dissipation) in both medium 1 and 2 \cite{BoardmanPRB1981,EconomouPR1969}, which demands $k_z$ lying outside the light
cone of the dielectric according to Eq.~(\ref{RadialWavenumber}). It turns out that the index
$\varsigma$ would have to be real-valued as well, to support the unique plasmonic edge modes \cite{DobrzynskiPRB1972,BoardmanPRB1981}.
Given a set of state parameters $\{\kappa_r, \varsigma, k_z\}$, the scalar potentials $\PhTM$ and
$\PhTE$ in a specific sector take the forms of
\begin{align}
\PhTM(r,\phi,z;\kappa_r,\varsigma,k_z) &= \frac{1}{\epsilon \kappa_r} \mathrm{K}_{\Ii\varsigma}(\kappa_r r) \left[ A_\varsigma
\Ee^{-\varsigma\phi} + B_\varsigma \Ee^{+\varsigma\phi} \right]\Ee^{\Ii k_z z}, \label{ElectricScalarPotential}\\
\PhTE(r,\phi,z;\kappa_r,\varsigma,k_z) &= \frac{1}{\mu \kappa_r} \mathrm{K}_{\Ii\varsigma}(\kappa_r r) \left[ C_\varsigma
\Ee^{-\varsigma\phi} + D_\varsigma \Ee^{+\varsigma\phi} \right]\Ee^{\Ii k_z z},
\label{MagneticScalarPotential}
\end{align}
in which we have omitted the time-harmonic factor $\Ee^{-\Ii\omega t}$ but keep
the relation Eq.~(\ref{RadialWavenumber}) in mind. $A_\varsigma$, $B_\varsigma$, $C_\varsigma$ and
$D_\varsigma$ are all undetermined coefficients. $\mathrm{K}_\nu$ is the $\nu$th-order modified
Bessel function of the second kind. This type of Bessel function guarantees convergence as
$r\rightarrow\infty$ for arbitrarily complex-valued orders ($\nu=\Ii\varsigma$) and arguments
($\kappa_rr=-\Ii k_rr$) \cite{AbramowitzMathFunctions1965}. In Fig.~\ref{FigKelvinFunction}, we plot
$\mathrm{K}_{\Ii\varsigma}(\kappa_r r)$ for real-valued
$\varsigma$ in both the $\kappa_rr$ scale and $\ln(\kappa_rr)$ scale. This function exhibits source-free indefinite oscillation at small argument ($\kappa_rr\rightarrow0$ or $\ln(\kappa_rr)\rightarrow-\infty$) but evanescent decay at large argument. This behavior is completely different from the function $\mathrm{K}_{\nu}(\kappa_r r)$ for real-valued $\nu$, which undergoes straight exponential decay (from potentially a line source at $r=0$) \cite{AbramowitzMathFunctions1965}. If measured in terms of the coordinate $r$, the oscillating and decaying regions of $\mathrm{K}_{\Ii\varsigma}(\kappa_r r)$ are separated approximately at
\begin{equation}
r \simeq \frac{\varsigma}{\kappa_r} \equiv b,\label{PropagatingRadius}
\end{equation}
in which $b$ defines a natural bounding radius. The waves are standing in the region $r \lesssim b$
and only weakly penetrating into the region $r \gtrsim b$. No matter how large $b$ is, these waves
are always radially bounded (non-radiative), even though the material itself is radially unbounded. As we
shall discuss in detail below, this is a hallmark of the plasmonic edge modes at deep axial
subwavelength in a metal-dielectric sectorial structure.

Implementing the boundary conditions for an arbitrary axial wavenumber $k_z$ in the systems
containing sharp wedges is mathematically challenging. Rigorous derivation requires the complicated
Kontorovich-Lebedev integral transform over the index $\varsigma$ \cite{RawlinsPRSLA1999}. To reveal the crucial physics most relevant to our interest, we will make use of the indefinite signature of metal-dielectric structures and particularly investigate the eigenmodes at deep axial subwavelength,
$k_z^2\gg|\mu\epsilon|\omega^2/c^2$, which take on minimal coupling with free-space photons. This allows for an asymptotically identical $\kappa_r$ in both media 1
and 2,
\begin{equation}
\kappa_r^2 \simeq k_z^2.
\label{NonretardedApproximation}
\end{equation}
In this scenario, the system is non-retarded in the $r\phi$-plane and the boundary connection
is greatly simplified. The terms explicitly carrying $\omega/ck_z$ in
Eqs.~(\ref{ElectricField}) and (\ref{MagneticField}) can be dropped, leading to the decoupled electrostatic modes with vanishing magnetic field and magnetostatic modes with vanishing electric field,
\begin{align}
\left[\partial_r^2+\frac{1}{r^2}\partial_\phi^2 - k_z^2 \right] \PhTM \simeq 0 , \qquad \bm{E}
\simeq -\nabla \PhTM, \qquad \bm{H} \simeq 0, \label{ElectrostaticPotential} \\
\left[\partial_r^2+\frac{1}{r^2}\partial_\phi^2 - k_z^2\right] \PhTE \simeq 0 , \qquad \bm{H}
\simeq -\nabla \PhTE, \qquad \bm{E} \simeq 0. \label{MagnetostaticPotential}
\end{align}
The basic solutions of $\PhTM$ and $\PhTE$ in a specific sector keep unchanged from
Eqs.~(\ref{ElectricScalarPotential}) and (\ref{MagneticScalarPotential}) except for the asymptotic
relation Eq.~(\ref{NonretardedApproximation}) replacing Eq.~(\ref{RadialWavenumber}).

In the following sections, we will focus on solving the plasmon-related electrostatic modes in our
metal-dielectric sectorial structure. It is important to clarify that although the electrostatic
approximation initially deduced from $k_z^2\gg|\mu\epsilon|\omega^2/c^2$ seems quite radical, it is
in fact a surprisingly good approximation that can work in a much wider range than expected.
Boardman \textit{et al.} \cite{BoardmanPRB1981} once calculated exactly the guided plasmonic edge modes on a single
parabolic metal wedge in vacuum, taking into account the retardation effect. They proved that the
electrostatic approximation, especially when the wedge is sharp, gives satisfactory results even if
the dispersion curves may have nearly touched the light line of the dielectric.

\section{Spectral analysis}

As shown in Fig.~\ref{FigSectorDiagram}, our structure is periodic in the $\phi$-direction. The
transfer matrix traversing one angular unit can be derived as
\begin{align}
\mathbf{T} &=
\begin{pmatrix}
\Ee^{-\varsigma\GmA}\left[\cosh(\varsigma\GmB) -
\frac{\EpsA^2+\EpsB^2}{2\EpsA\EpsB}\sinh(\varsigma\GmB)
\right] & \frac{\EpsA^2-\EpsB^2}{2\EpsA\EpsB}\sinh(\varsigma\GmB) \\
\frac{\EpsB^2-\EpsA^2}{2\EpsA\EpsB}\sinh(\varsigma\GmB)  & \Ee^{+\varsigma\GmA}
\left[\cosh(\varsigma\GmB) +
\frac{\EpsA^2+\EpsB^2}{2\EpsA\EpsB}\sinh(\varsigma\GmB)
\right] \\
\end{pmatrix}.
\end{align}
The eigen-spectrum can be solved in view of the Bloch theorem \cite{FerrandoPRE2005,AshcroftSolidState1976},
\begin{align}
\det \left|\mathbf{T} - \Ee^{\Ii h \gamma }\mathbf{I}\right| =0, \quad \left(h = 0, \pm1, \pm2, \dots,
\pm \frac{N}{2}\right).\label{DiscreteMomentum}
\end{align}
Recall $\gamma=2\pi/N$, where $\gamma=\gamma_1+\gamma_2$ is the angular periodicity, $N$
is the total unit number. We can obtain an elegant ``band" equation akin to
that of the Kronig-Penney model in solid state physics \cite{AshcroftSolidState1976} (but winded into a $2\pi$ circle here),
\begin{align}
\cos \left(h \gamma \right) &= \cosh(\varsigma\GmA)\cosh(\varsigma\GmB) +
\frac{1}{2}\left(\frac{\EpsA}{\EpsB}+\frac{\EpsB}{\EpsA}\right)\sinh(\varsigma\GmA)\sinh(\varsigma\GmB).
\label{ElectricBandEquation}
\end{align}
The azimuthal wavenumber $h$ denotes the structure-modulated angular momentum about the $z$-axis, whose
upper limit is at the boundary of the first angular Brillouin zone $\pm N/2$ determined from
material design. In the continuous limit $N\rightarrow\infty$, $\gamma\rightarrow0$, $h$
approaches the $z$-component of the true angular momentum $J_z$ of the plasmon-polaritons in this
structure, and can take however large values. If we perform a series expansion to $h\gamma$,
$\varsigma\GmA$ and $\varsigma\GmB$ in Eq.~(\ref{ElectricBandEquation}) under the continuous limit, we can find a quite appealing result,
\begin{align}
\frac{\varsigma^2}{\EffEpsPh} + \frac{h^2}{\EffEpsRd} = 0, \label{ApproxBandEquations}
\end{align}
where the effective permittivities from the effective medium theory automatically show up \cite{SmolyaninovPRL2010,JacobOE2006},
\begin{align}
\EffEpsRd = \EpsA \EtaA + \EpsB \EtaB ,\quad \EffEpsPh = \frac{\EpsA\EpsB}{\EpsA \EtaB + \EpsB
\EtaA},\label{EffectiveParameters}
\end{align}
in which $\EtaA\equiv\GmA/\gamma$ and $\EtaB\equiv\GmB/\gamma$ are the filling ratios of medium 1
and 2, respectively. As can be imagined, if $\EffEpsRd$ and $\EffEpsPh$ are of opposite signs,
Eq.~(\ref{ApproxBandEquations}) clearly demonstrates the indefinite signature of this metamaterial,
which possesses singular density of states on iso-frequency surfaces \cite{SmolyaninovPRL2010,KrishnamoorthyScience2012}. The right-hand side of
Eq.~(\ref{ApproxBandEquations}) does not have a usual term like $\omega^2/c^2$ because of the
non-retarded regime (equivalently the $c\rightarrow \infty$ limit) that we have chosen; however,
the nontrivial frequency dependence is still implicitly enclosed in $\EffEpsRd$ and $\EffEpsPh$.

\begin{figure}[htbp]
\centerline{\includegraphics[scale=0.75]{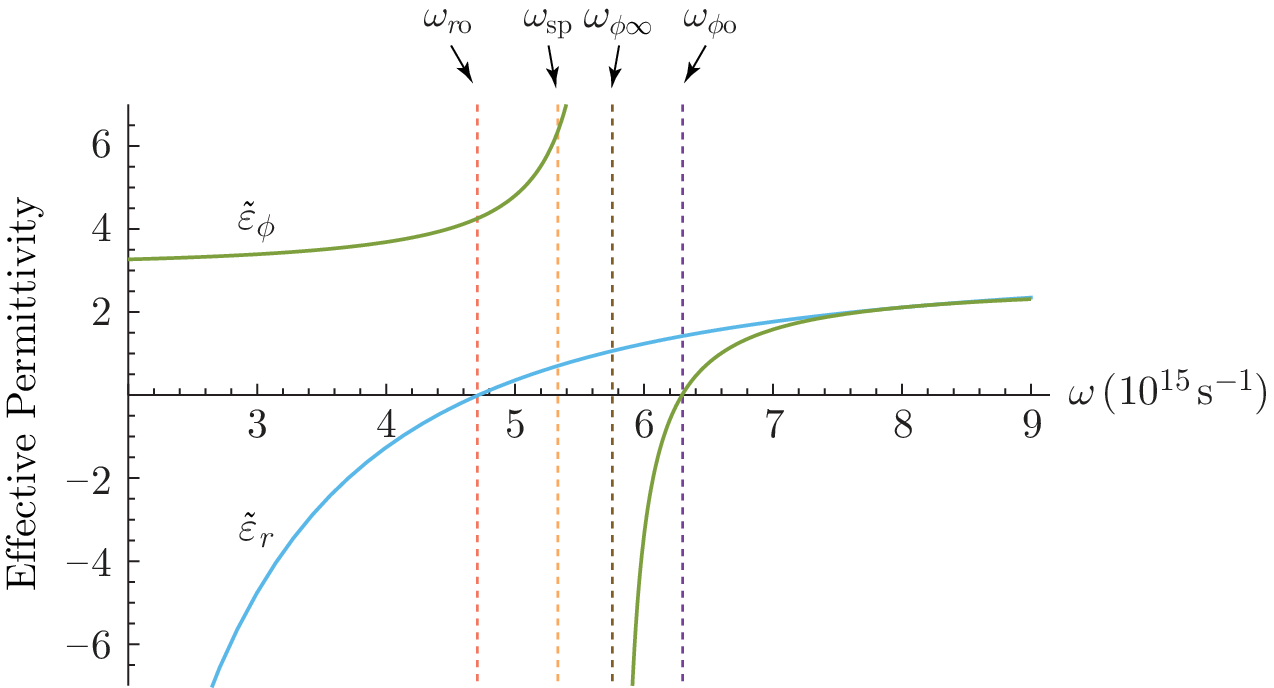}} \caption{Effective permittivities
$\EffEpsRd$ and $\EffEpsPh$ versus frequency $\omega$ for the metal and dielectric filling ratios
$\EtaA=\frac{1}{3}$ and $\EtaB=\frac{2}{3}$.}\label{FigEffectivePermittivity}
\end{figure}

Hereafter, we study the experimentally accessible metal-dielectric construction. We choose silver
(Ag) as medium 1 and silicon dioxide (SiO$_2$) as medium 2. Their permeabilities $\MuA$ and $\MuB$
are set to be 1. Their permittivities in the 200~--~2000~nm wavelength range can be well fitted by
a frequency-dependent modified Drude model \cite{KikPRB2004},
\begin{equation}
\EpsA(\omega) \approx \epsilon_{\mathrm{h}} - (\epsilon_{\mathrm{s}}-\epsilon_{\mathrm{h}})
\frac{\omega_{\mathrm{p}}^2}{\omega^2-\Ii\omega \Gamma},
\end{equation}
where $\epsilon_{\mathrm{h}}=5.45$, $\epsilon_{\mathrm{s}}=6.18$,
$\omega_{\mathrm{p}}=17.2\times10^{15}$~s$^{-1}$, $\Gamma=8.35\times10^{13}$~s$^{-1}$, and a nearly
frequency-independent constant $\EpsB\approx2.13$. For the proof-of-concept analysis here,
we neglect the dissipation rate $\Gamma$ of silver. As we know, for example, silver nanowires in
SiO$_2$ have a propagation length of at least several microns even if the wire radius may be
smaller than 50~nm and the operating wavelength may be shorter than 500~nm \cite{MAOA2010,LiOE2013}. Although our
sectorial structure is different from the circular wires, fundamentally they share a similar dissipation length scale. Let us now take a look at the
behavior of effective permittivities $\EffEpsRd$ and $\EffEpsPh$ for some given metal and
dielectric filling ratios. We use $\EtaA=\frac{1}{3}$ and $\EtaB=\frac{2}{3}$ as an example
throughout this paper. Figure \ref{FigEffectivePermittivity} shows the change of $\EffEpsRd$ and
$\EffEpsPh$ versus frequency. As can be seen, there exist several characteristic frequencies:
$\omega_{r\mathrm{\SSS o}} = 4.72\times10^{15}$~s$^{-1}$ is the frequency as $\EffEpsRd(\omega)=0$;
$\omega_{\phi\mathrm{\SSS o}} = 6.29\times10^{15}$~s$^{-1}$ is the frequency as
$\EffEpsPh(\omega)=0$; $\omega_{\phi\infty} = 5.76\times10^{15}$~s$^{-1}$ is the frequency as
$\EffEpsPh(\omega)=\infty$. Accordingly, $\EffEpsRd$ and $\EffEpsPh$ change signs in the different
frequency ranges divided by these characteristic frequencies. There is an another characteristic
frequency, i.e., the metal-dielectric surface plasma frequency, $\omega_{\mathrm{sp}} =
\omega_{\mathrm{p}}\sqrt{(\epsilon_{\mathrm{s}}-\epsilon_{\mathrm{h}})/(\epsilon_{\mathrm{h}}+\EpsB)}
=5.34\times10^{15}$~s$^{-1}$, between $\omega_{r\mathrm{\SSS o}}$ and $\omega_{\phi\infty}$. These
frequencies will be frequently referred to below.

\begin{figure}[htbp]
\centerline{\includegraphics[scale=0.75]{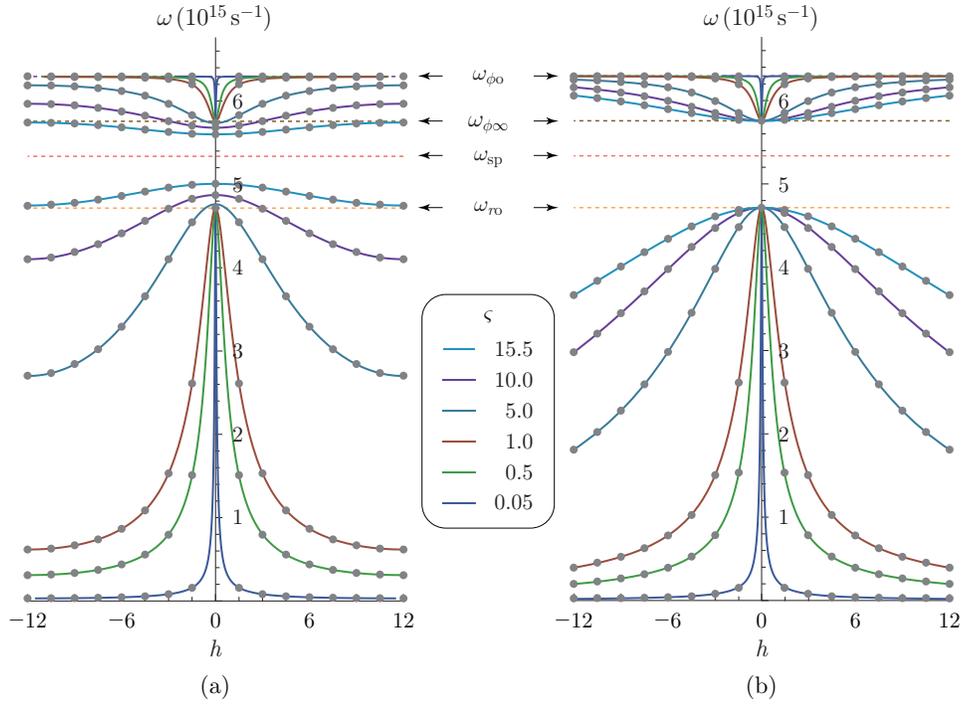}} \caption{Calculated eigen-spectrum
$\omega(\varsigma,h)$ versus $h$ for several fixed $\varsigma$. (a) From the actual medium theory
with $N=24$, $\gamma=\pi/12$, and $h$ cutoff at the 1st Brillouin zone boundary
$\pm\pi/\gamma=\pm12$. (b) From the effective medium theory with $h$ manually cutoff at $\pm12$ for
comparison purposes. Note that a physical $h$ can only take discrete integers denoted by grey dots according to Eq.~(\ref{DiscreteMomentum}).}\label{FigBandStructure}
\end{figure}

In Fig.~\ref{FigBandStructure}, we plot the eigen-spectrum $\omega(\varsigma,h)$ versus $h$ for
several fixed $\varsigma$ based on the actual medium theory Eq.~(\ref{ElectricBandEquation}) and
the effective medium theory Eq.~(\ref{ApproxBandEquations}), respectively. $\varsigma$ resembles a
band index in the band theory of electrons in solids. For the actual spectrum shown in
Fig.~\ref{FigBandStructure}(a), we choose $N=24$, $\gamma=\pi/12$, as an example; the structure-modulated
angular momentum $h$ is limited within the 1st Brillouin zone $|h|<\pi/\gamma=N/2=12$. For the
effective spectrum shown in Fig.~\ref{FigBandStructure}(b), these restrictions are irrelevant (or
strictly speaking, $N=\infty$, $\gamma=0$); but for comparison purposes, we only draw
$|h|<\pi/\gamma=12$ in the same angular momentum range. One should keep in mind that $h$ can only
take discrete integers according to Eq.~(\ref{DiscreteMomentum}), meaning that the continuous curves in the graphs should be regarded as the connecting curves for the grey dots. For every given $\varsigma$, there are always lower-energy and higher-energy two
$\omega(\varsigma,h)$ curves in the regions $\omega<\omega_{\mathrm{sp}}$ and
$\omega_{\mathrm{sp}}<\omega<\omega_{\phi\mathrm{\SSS o}}$, respectively. The main qualitative
difference between Figs.~\ref{FigBandStructure}(a) and \ref{FigBandStructure}(b) is that (b) shows
an energy gap in the frequency range $\omega_{r\mathrm{\SSS o}}<\omega<\omega_{\phi\infty}$ for
however large $\varsigma$, whereas (a) permits the large-$\varsigma$ curves to penetrate into the gap region and approach the $\omega=\omega_{\mathrm{sp}}$ line from two sides. The reason for the
existence of an energy gap in the effective medium theory can be intuitively grasped from
Fig.~\ref{FigEffectivePermittivity}, in which the frequency range $\omega_{r\mathrm{\SSS
o}}<\omega<\omega_{\phi\infty}$ (and also $\omega>\omega_{\phi\mathrm{\SSS o}}$) embodies both
positive $\EffEpsRd$ and positive $\EffEpsPh$. This enforces $\varsigma$ to be imaginary-valued
(refer to Eq.~(\ref{ApproxBandEquations})) and so kills the plasmonic edge modes in our structure.
However, the actual medium theory from Eq.~(\ref{ElectricBandEquation}) implies that this picture
is only approximately correct. If the band index $\varsigma$ that controls the azimuthal
confinement and the radial oscillation (for a given $\kappa_r$) is extraordinarily large, the
effective medium theory naturally breaks down, and the surface plasmons on different wedge
interfaces fully decouple from each other and converge independently towards the same extreme
short-wavelength limit $\omega=\omega_{\mathrm{sp}}$. Furthermore, even for the small-$\varsigma$
curves, Figs.~\ref{FigBandStructure}(a) and \ref{FigBandStructure}(b) still show prominent
frequency difference when $h$ approaches the Brillouin zone boundary.
Only in the region where both
$\varsigma$ and $h$ are small, the two formalisms agree well with each other. All the curves in
this region pass through either the $h=0, \omega=\omega_{r{\SSS\mathrm{o}}}$ point or the $h=0,
\omega=\omega_{\phi\infty}$ point, very insensitive to the value of $\varsigma$. This property can
be deduced from Eq.~(\ref{ApproxBandEquations}) supposing either $\EffEpsRd\simeq0$ or
$\EffEpsPh\simeq\infty$, which is an index-near-zero (INZ) or index-near-infinity (INI) behavior.

The aforementioned eigen-spectrum $\omega(\varsigma,h)$ is independent of the axial wavenumber $k_z$, which seems
unusual from the perspective of waveguide theory. This is a result of both the deep-subwavelength limit and the
perfect wedge tips that we have assumed at $r=0$. As demonstrated below, once we introduce
rounded wedges, even if we still keep the deep-subwavelength limit, $\omega(\varsigma,h)$ will become $k_z$-dependent.

\section{Rounded wedges}

The electrostatic potential of the calculated modes above oscillates indefinitely at $r=0$, which can be inferred
from the asymptotic behavior of $\mathrm{K}_{\Ii\varsigma}(\kappa_rr)$ at small argument (see Fig.~\ref{FigKelvinFunction}) \cite{AbramowitzMathFunctions1965,DavisPRB1976},
\begin{equation}
\mathrm{K}_{\Ii\varsigma}(\kappa_rr) \sim -\sqrt{\frac{\pi}{\varsigma\sinh\varsigma}}
\sin\left[\varsigma\ln\left(\frac{1}{2}\kappa_rr\right)-\arg\Gamma(1+\Ii\varsigma)\right],\quad
(\kappa_r r\rightarrow 0),
\end{equation}
where $\arg\Gamma$ is the complex phase angle of gamma function. In addition, the radial and
azimuthal components of the field diverge like $1/\kappa_rr$ as $r\rightarrow 0$ (refer to
Eq.~(\ref{ElectrostaticPotential})). These singular behaviors are due to the infinite charge
accumulation at the infinitely sharp tips. While the strong field enhancement at the structural center is physical and is favorable for nanophotonics, the mathematical artifacts must be removed from the theory. Technically, the wedges are always rounded and can
never seamlessly touch each other under fabrication. To make our theoretical study match better
with the reality, we adopt a conformal coordinate mapping to conveniently achieve the rounded and
gapped configurations, which automatically removes the divergence and indefinite oscillation, and
can reveal more subtle physics around the wedge tips \cite{BoardmanPRB1981,DavisPRB1976}.

\begin{figure}[htbp]
\centerline{\includegraphics[scale=0.75]{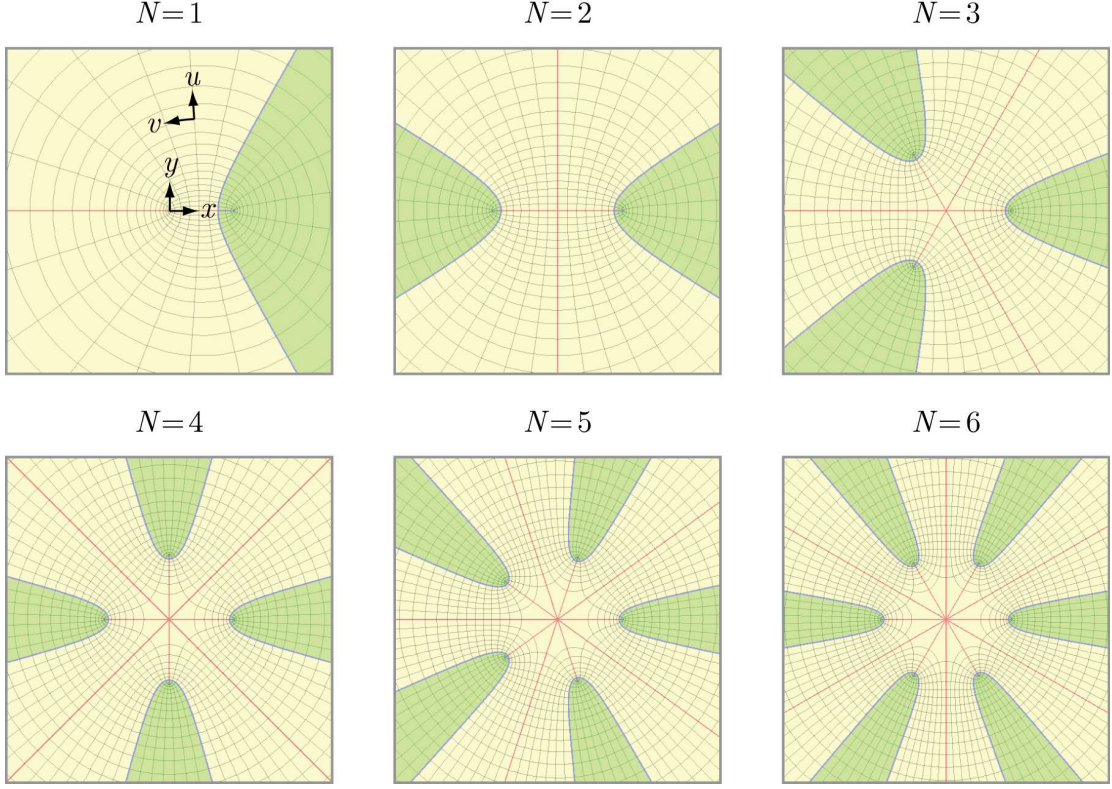}} \caption{Conformal mapping from the
$(x,y)$-coordinates to the $(u,v)$-coordinates with different unit numbers $N=1$ to 6,
respectively. The red lines show the branch cuts and semi-focal length. The green areas are the intended areas for metal filling at the ratio
$\EtaA=\frac{1}{3}$.}\label{FigConformalMapping}
\end{figure}

Let us momentarily write our electrostatic scalar potential in the $(x,y)$-coordinates and again
omit the phase factor $\Ee^{\Ii k_zz}\Ee^{-\Ii\omega t}$,
\begin{equation}
\left[\partial_x^2+\partial_y^2 - k_z^2 \right]\PhTM(x,y) = 0.
\end{equation}
First, we define two sets of (dimensionless) complex coordinates $w=(x+\Ii y)/a$ and $s=u+\Ii v$,
where $a$ is for now a characteristic length parameter that cancels the dimension of $x$ and $y$.
Next, we connect the two coordinate systems by a conformal mapping $w = \Lambda(s)$, where
$\Lambda$ is an analytical function. Thus the Helmholtz equation in the $(u,v)$-coordinates reads
\begin{equation}
\left[\partial_u^2+\partial_v^2 - \left|\frac{\Dd w}{\Dd s}\right|^2_{(u,v)} k_z^2a^2
\right]\PhTM(u,v) = 0. \label{OldHelmholtzEquation}
\end{equation}
We introduce the following conformal mapping for any desired unit number $N=1, 2, 3,\dots$,
\begin{equation}
\begin{split}
w^N = \left[\cosh s\right]^2, \quad\text{ i.e. },\quad x+\Ii y = a\left[\cosh
(u+\Ii v)\right]^{\frac{2}{N}}\Ee^{\Ii \frac{2\pi}{N}n}, \\
\left(u\in[0,+\infty),\quad v\in \left[-\frac{\pi}{2},+\frac{\pi}{2}\right], \quad n=0,1,2,\dots,N-1\right).
\end{split}
\end{equation}
The new $(u,v)$-coordinate system constitutes a generalized elliptic cylinder coordinate system, with $N$ sectors partitioned by $N-1$ branch cuts. The index $n$ denotes which sector a
$(x,y)$-point maps in. The length parameter $a$ is the semi-focal length measured in the old
$(x,y)$-coordinate system. Far from the coordinate center ($u\gg 1$), we have
\begin{equation}
x+\Ii y = r\Ee^{\Ii \phi} \simeq \frac{a}{\sqrt[N]{4}}\Ee^{\frac{2}{N}u} \Ee^{\Ii
\left(\frac{2}{N}v+\frac{2\pi}{N}n\right)},\quad\text{ i.e. },\quad r \simeq
\frac{a}{\sqrt[N]{4}}\Ee^{\frac{2}{N}u}, \quad \phi \simeq \frac{2}{N}v+\frac{2\pi}{N}n,
\label{FarCoordinates}
\end{equation}
where $u$ resembles the logarithm of the radial coordinate $r$ while $v$ (together with $n$)
resembles the azimuthal coordinate $\phi$ in spite of some proportionality constants.
Figure \ref{FigConformalMapping} shows the new $(u,v)$-coordinate lines for $N=1$ to 6. As can be
seen, the conformal coordinate mapping automatically preserves the orthogonality. The shaded areas
in Fig.~\ref{FigConformalMapping} are to be filled with metal at the same filling ratio $\EtaA
= \frac{1}{3}$ as in the preceding part, straightforwardly corresponding to the regions with
$0\leq u<+\infty$, $-\frac{\pi}{2}\EtaA<v<+\frac{\pi}{2}\EtaA$, $n=0,1,2,\dots,N-1$. All the metal
wedges near the center are naturally rounded following the coordinate lines of $u$. The unshaded
areas in Fig.~\ref{FigConformalMapping} are to be filled with dielectric. These configurations
nicely imitate the actual structures produced through nanofabrication, and are much more realistic
than the one illustrated in Fig.~\ref{FigSectorDiagram}.

The transformation function appearing in Eq.~(\ref{OldHelmholtzEquation}) is
\begin{equation}
\left|\frac{\Dd w}{\Dd s}\right|^2 = \left|\frac{2}{N}\left[\cosh
s\right]^{\frac{2}{N}-1} \sinh s\right|^2 =
\frac{4}{N^2}\left[\sinh^2u+\cos^2v\right]^{\frac{2}{N}-1}
\left[\sinh^2u+\sin^2v\right].
\end{equation}
Therefore, the new Helmholtz equation looks like a Schr\"odinger equation in a strange potential,
\begin{equation}
-\left[\partial_u^2+\partial_v^2\right]\PhTM +
\frac{4}{N^2}k_z^2a^2\left[\sinh^2u+\cos^2v\right]^{\frac{2}{N}-1}
\left[\sinh^2u+\sin^2v\right]\PhTM = 0.\label{NewHelmholtzEquation}
\end{equation}
For $N=1$ and $N=2$, the partial differential equation is separable, which has solutions in the
form of Mathieu functions \cite{McLachlanMathieuFunctions1951}. But for $N\geq3$, the equation is only approximately separable;
for instance, in the region close to the wedge tips $u\simeq0$, $v\simeq0$ \cite{DavisPRB1976}, the major behaviors can
be described by a quadratic expansion to the $\sinh$, $\sin$ and $\cos$ functions,
\begin{equation}
-\left[\partial_u^2+\partial_v^2\right]\PhTM + \left(\frac{2}{N}k_za\right)^2 \left(u^2+v^2\right)
\PhTM = 0,\label{QuadraticExpansion}
\end{equation}
This leads to the separable eigen-equations for $\PhTM(u,v)\equiv U(u)V(v)$ with an eigenvalue
$\left(\frac{2}{N}\varsigma\right)^2$,
\begin{align}
\left[-\frac{\Dd^2}{\Dd u^2} + \left(\frac{2}{N}k_za\right)^2 u^2 \right] U(u) &= +\left(\frac{2}{N}\varsigma\right)^2 U(u),\label{UEquationHarmonic}\\
\left[-\frac{\Dd^2}{\Dd v^2} + \left(\frac{2}{N}k_za\right)^2 v^2 \right] V(v) &=
-\left(\frac{2}{N}\varsigma\right)^2 V(v).\label{VEquationHarmonic}
\end{align}
Referring to Eq.~(\ref{FarCoordinates}) we shall see that as $u\gg 1$ the index $\varsigma$
introduced in this way is identical to the $\varsigma$ that we used earlier.

\begin{figure}[htbp]
\centerline{\includegraphics[scale=0.75]{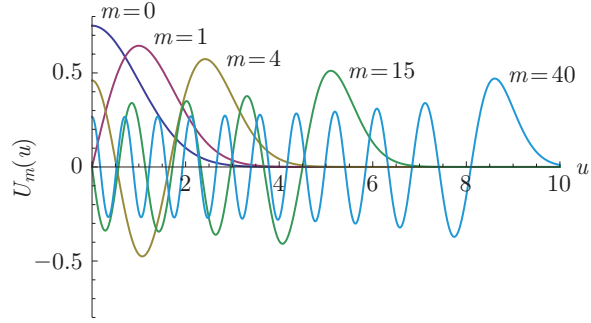}} \caption{Plots of the normalized Hermite function with $m=0, 1, 4, 15, 40$.}\label{FigHermiteFunction}
\end{figure}

The radial part of the problem around the wedge tips coincides with the quantum harmonic oscillator problem. The eigen-solutions are bounded in $u\in[0,+\infty)$ with the discrete eigenvalues denoted by an integer $m$ \cite{AbramowitzMathFunctions1965,DavisPRB1976},
\begin{align}
U_m(u) &= \frac{1}{\sqrt{2^m m!}}\left(\frac{2k_za}{ N\pi }\right)^{\frac{1}{4}}\exp \left[-\frac{1}{2}\left(\frac{2}{N}k_za\right) u^2\right] \label{USolution}
\mathcal{H}_m \left[\left(\frac{2}{N}k_za\right)^{\frac{1}{2}}u\right],\\
\varsigma_m &= \sqrt{\frac{N k_za}{2}(2m+1)},\quad (m=0,1,2,3,\dots), \label{DiscreteIndex}
\end{align}
where $\mathcal{H}_m$ is the $m$th-order Hermite polynomial. For clarity, in Fig.~\ref{FigHermiteFunction} we plot the normalized Hermite function for a few integer $m$, which is just our radial solution $U_m(u)$ in Eq.~(\ref{USolution}) taking $2k_za/N=1$. As can be surmised, the lower the order $m$ is, the more trapped the plasmonic edge modes lie around the metal wedge tips $u\simeq0$. Interestingly, Fig.~\ref{FigHermiteFunction} and Fig.~\ref{FigKelvinFunction}(b) look quite alike each other, noting that $u\sim\ln (r/a)$ from Eq.~(\ref{FarCoordinates}). Indeed, $\varsigma_m$ (or simply $m$) controls the (finite) number of radial oscillation and the bounding radius around the structural center, just as what $\varsigma$ does in the far region. They asymptotically merge with each other.

The azimuthal part of the solution takes linear combinations
of the parabolic cylinder functions \cite{AbramowitzMathFunctions1965,DavisPRB1976},
\begin{align}
V_m(v) & = A_m\mathcal{D}_{-m-1}\left[+\left(\frac{4}{N}k_za\right)^{\frac{1}{2}}v\right] + B_m
\mathcal{D}_{-m-1}\left[-\left(\frac{4}{N}k_za\right)^{\frac{1}{2}}v\right],\label{VSolution}
\end{align}
in which the first and second term looks somewhat like $\Ee^{-\frac{2}{N}\varsigma v}$ and
$\Ee^{+\frac{2}{N}\varsigma v}$, respectively, when $v$ is small. Applying the continuity
conditions of $V(v)$ and $\epsilon\partial_vV(v)$ across the metal-dielectric interfaces at
$v=-\frac{\pi}{2}\EtaA$ and $+\frac{\pi}{2}\EtaA$, and utilizing again the Bloch theorem, a ``band"
equation conceptually equivalent to but algebraically more
complex than Eq.~(\ref{ElectricBandEquation}) can be derived. For the sake of concision, we do not present it here. The main conclusion is that the new eigen-spectrum $\omega(k_za,m,h)$ after considering the rounded wedges is discretized by $m$ and dependent on $k_za$ through Eq.~(\ref{DiscreteIndex}). $h$ is still the structure-modulated angular momentum.

\begin{figure}[htbp]
\centerline{\includegraphics[scale=0.75]{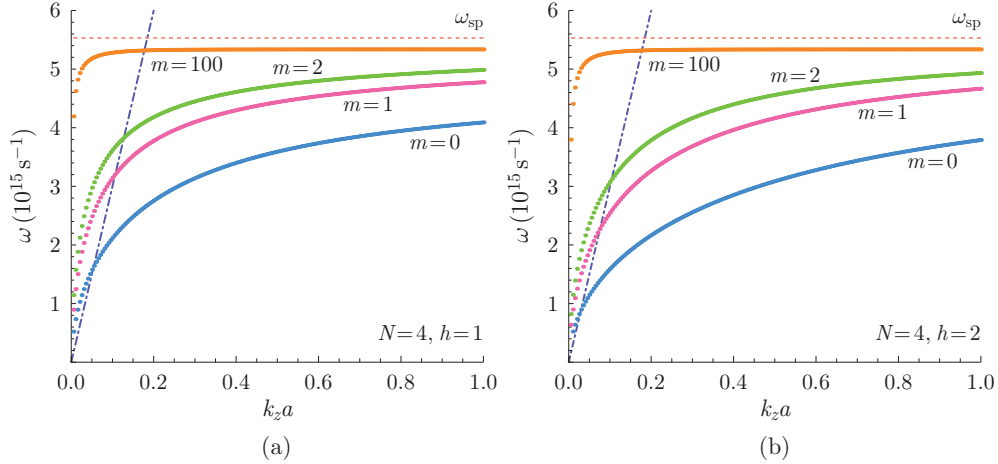}} \caption{Calculated eigen-spectrum
$\omega(k_za,m,h)$ versus $k_za$ in the case of $N=4$. (a) $h=1$, $m=0$, $1$, $2$, $100$. (b)
$h=2$, $m=0$, $1$, $2$, $100$. The dot-dashed lines are the light line of the dielectric taking $k_za$ as the abscissa when $a=10$~nm. As a result of the non-retarded approximation, only the right portion of the calculated dispersion
curves outside the light cone is physical. The dotted lines are the limiting dispersion curves as
$m\rightarrow\infty$, converging to the surface plasma frequency
$\omega_{\mathrm{sp}}$.}\label{FigEdgeSpectrum}
\end{figure}

In Fig.~\ref{FigEdgeSpectrum}, we plot $\omega(k_za,m,h)$ as a function of $k_za$ for several given $m$ and $h$. The material model used for $\EpsA$ of silver and $\EpsB$ of silicon dioxide is the same as before. In the non-retarded regime, the light speed $c$ does not enter our theory and so the semi-focal length $a$ becomes the only length parameter of the system. At this stage, we choose $a=10$~nm in accordance with the typical linewidth of today's nano-lithography. The dispersion curves from our calculation intersect the light line, but in reality they may bend more quickly to zero before touching the light line \cite{BoardmanPRB1981,EconomouPR1969}. Even though our non-retarded approximation cannot reproduce such a feature, the right portion of the curves outside the light cone is reliable \cite{BoardmanPRB1981}. At a given $k_za$ and $h$, the larger the $m$ is, the more radial oscillation there is, and the higher the needed frequency $\omega$ is. In general, larger-$m$ curves appear flatter, meaning a less $k_za$-dependence, which is qualitatively consistent with the vanishing $k_z$-dependence in the $\omega(\varsigma,h)$ curves that we have obtained earlier without considering the tip effect.  In the extreme case $m\rightarrow\infty$ here (by analogy with the $\varsigma\rightarrow\infty$ case in Fig.~\ref{FigBandStructure}), all the dispersion curves are pushed towards the line of surface plasma frequency $\omega_{\mathrm{sp}}$. Comparing between Figs.~\ref{FigEdgeSpectrum}(a) and \ref{FigEdgeSpectrum}(b), we shall notice that for a fixed unit number $N$, a higher angular momentum $h$ tends to drag all the curves downwards and so permits exciting these modes at lower frequencies (but still large $k_z$). Likewise, we have also found that for a fixed $h$, a larger $N$ drags all the curves downwards and so permits lower-frequency excitation too. These higher $h$ or larger $N$ modes may be termed as the ``dark" modes in literature, i.e., the modes essentially prohibited from coupling with free-space light due to more profound symmetry-related reasons \cite{BenistyJOSAB2009}. For these modes, our non-retarded calculation is almost an exact calculation.

To be cautious, we should remember that the quadratic expansion adopted in Eq.~(\ref{QuadraticExpansion}) is quantitatively correct only for a relatively small $m$ and the near-tip region $u\sim 0$. For a large $m$ and the $u\gg1$ far region, the more accurate description is the continuous $\varsigma$ description that we have elaborated in the prior section; the small rounded wedges tips should not induce a sizable impact there.

\section{Field profiles}

After finding the eigen-solutions of the structure for any unit number $N$ and angular momentum $h$, we can obtain the coefficients $A_\varsigma$ and $B_\varsigma$ in Eq.~(\ref{ElectricScalarPotential}) or $A_m$ and $B_m$ in Eq.~(\ref{VSolution}) in all sectors. We
are then able to plot the field profiles, say, on the $z=0$ plane, for any wanted eigenmodes without resorting to the effective medium theory or numerical simulation, which is either invalid or inaccurate for structures of small sector numbers and sharp wedge tips.

\begin{figure}[htbp]
\centerline{\includegraphics[scale=0.75]{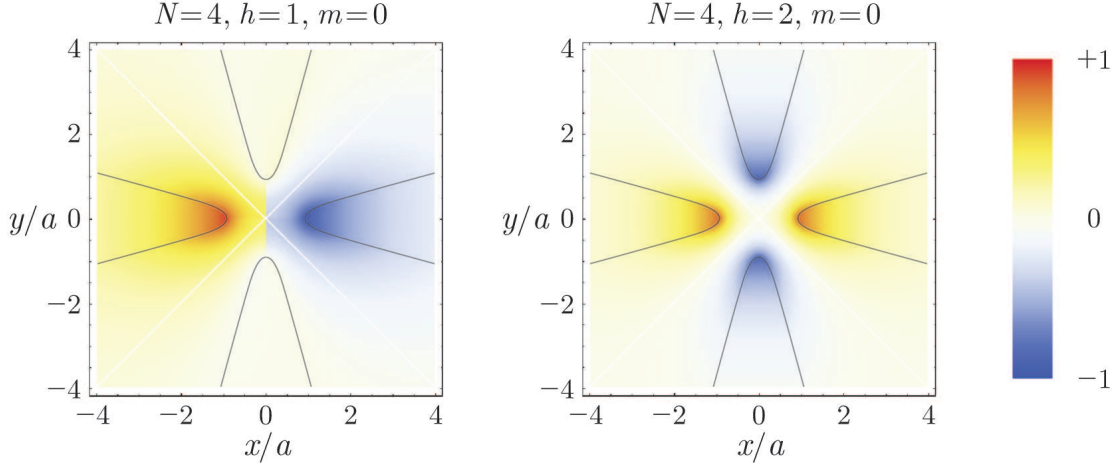}} \caption{Profiles of the electrostatic
potential close to the wedge tips at a representative frequency $\omega=3.54\times10^{15}$~s$^{-1}$
(532~nm free-space wavelength) for the unit number $N=4$, the structure-modulated angular momentum $h=1$ and
$2$, and the radial oscillation order $m=0$. The semi-focal length $a=10$~nm. The black curves indicate the wedge shapes. The white lines indicate the branch cuts from the conformal mapping. }\label{FigEdgeProfile}
\end{figure}

The finite semi-focal length $a$ in our conformal mapping physically designates a finite gap size
and tip radius of the rounded wedges, and hence settles a cutoff length in our problem. For whatever specified $\omega$ and $k_za$, the field strength cannot oscillate at an arbitrarily small spacing and must remain finite everywhere. As we have proved, the profiles of potential field $\PhTM$ can be approximated by the $m$th order Hermite polynomials close to the wedge tips while asymptotically approach the $\Ii\varsigma$th imaginary-order modified Bessel functions away from the tips. Let us first look at the profiles around the tips for some low-$m$ modes. If we choose $N=4$, $a=10$~nm, and $\omega=3.54\times10^{15}$~s$^{-1}$ (532~nm free-space wavelength), then according to Fig.~\ref{FigEdgeSpectrum} we realize that only the $m=0$ curves have an intersection point with $\omega=3.54\times10^{15}$~s$^{-1}$ at deep subwavelength, where $k_za=0.47$ for $h=1$, and $k_za=0.77$ for $h=2$. Fig.~\ref{FigEdgeProfile} displays the corresponding potential field profiles. The $h=1$ angular-momentum mode is of the cylindrical dipolar type with one sign change in the $2\pi$ azimuthal circle, and the $h=2$ angular-momentum mode is of the cylindrical quadrupolar type with two sign changes. The electric field $\bm{E}=-\nabla \PhTM$ as the derivative field of $\PhTM$ gains an enormous strength in the nanoscale tip regions.

\begin{figure}[htbp]
\centerline{\includegraphics[scale=0.75]{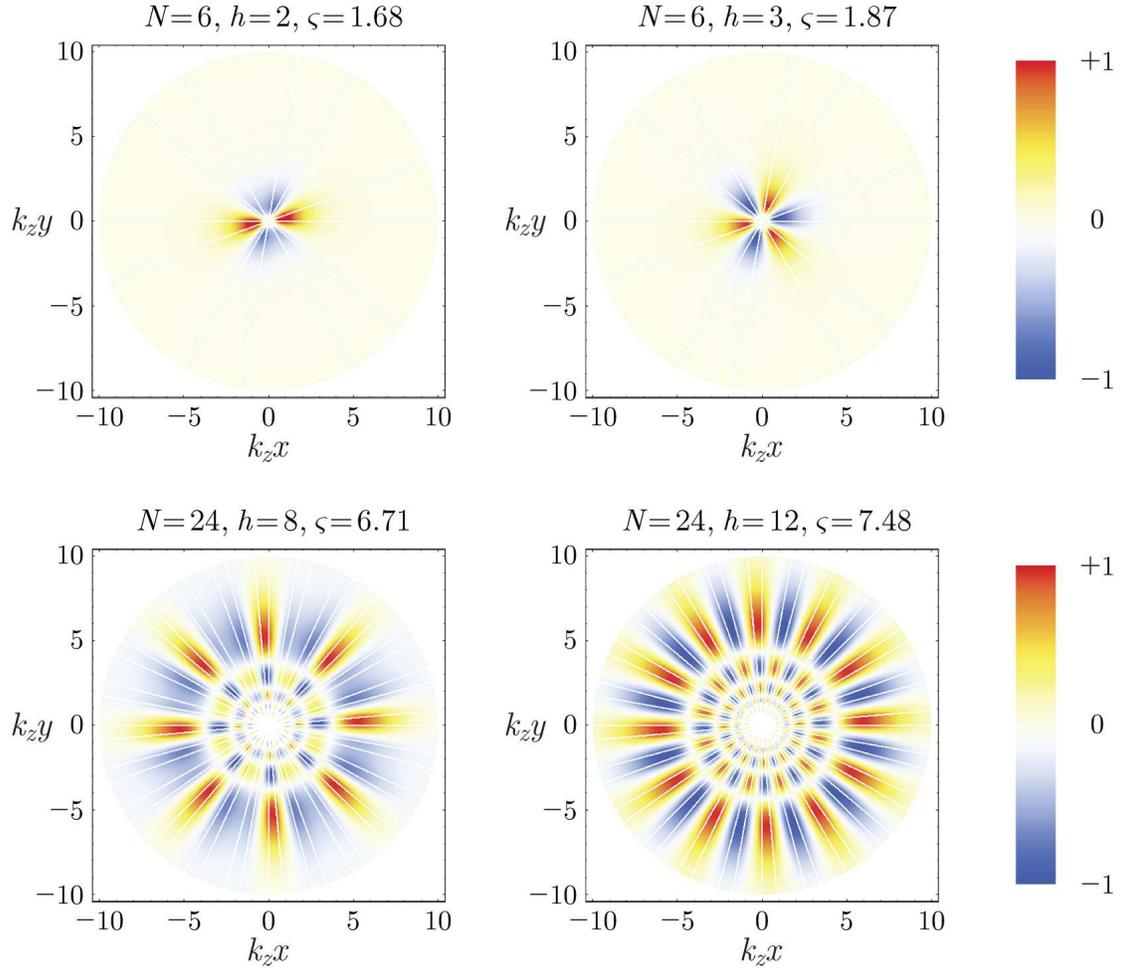}} \caption{Profiles of the electrostatic potential away from the wedge tips at a representative frequency
$\omega=3.54\times10^{15}$~s$^{-1}$ (532~nm free-space wavelength) for the unit numbers
$N=6$ and $24$, and several structure-modulated angular momentum $h$ chosen between 1 and $N/2$. The coordinates are measured in $k_zx$ and $k_zy$ in view of the non-retarded assumption $\kappa_rr\simeq k_zr$. The white lines indicate the wedge interfaces. The blank central areas are where the severe oscillation and field intensity occur.}\label{FigAngularFieldProfile}
\end{figure}

We next look at the field profiles slightly away from the tips (for example, $r>5a$, i.e., if $a\simeq10$~nm, $r\gtrsim 50$~nm). Figure \ref{FigAngularFieldProfile} displays the $N=6$ and $24$ two cases at several allowed $h$. In the far-from-tip region and deep-subwavelength regime, the eigen-spectrum and field profiles no longer depend on $k_z$ or $a$. Therefore, these plots are universal for any given $k_z$. We know from Eq.~(\ref{ElectricBandEquation}) that for the same $\omega$, an increasing $h$ increases the index $\varsigma$. Then if $k_z$ is fixed as well, a larger $\varsigma$ produces a larger bounding radius $b=\varsigma/k_z$. In agreement with our earlier argument, although the field prefers to localize closer to the structural center for a smaller $\varsigma$, it oscillates more drastically and spreads more widely away from the center for a larger $\varsigma$. But in any event, beyond the bounding radius, the field always quickly fades away. For a realistic material design using today's nano-lithography, $N$ perhaps cannot yet reach 24 that we have demonstrated. If we take the sub-wavelength $k_z$ to be of the order of $0.1$~nm$^{-1}$, and substitute in the calculated $\varsigma$ as labeled in Fig.~\ref{FigAngularFieldProfile}, $b$ is at most of the order of 100~nm, which means high plasmonic angular momentum can indeed be tightly trapped in an impressively narrow region in this structure.

In comparison, conventional dielectric waveguides, metallic nanowires, or metal-dielectric multilayer waveguides or cavities, cannot give rise to such a feature \cite{YehWaveguide2008,HuEPL2012,LiOE2013}. One may recall that the angular momentum modes in ordinary structures are described by the Bessel and Neumann functions (or the Hankel functions of the first and second kinds) with real-valued angular momentum index $\nu$ \cite{AbramowitzMathFunctions1965}. These functions either strongly converge to zero inside a diffraction-limited core when $\nu$ goes high, or have to be supported by a line source at $r=0$, and so do not represent the intrinsic modes of the systems \cite{JacobOE2006}. Metallic nanowires can to some degree support high plasmonic angular momentum at deep subwavelength. However, reducing the wire diameter to enhance the field intensity and confine the angular momentum in tens of nanometers is technically difficult.

\section{Conclusion}

We have performed a systematic study to the structure-modulated plasmonic angular momentum trapped at the center of a sectorial indefinite metamaterial. We have shown that the electric field associated with these angular momentum states is extremely intense in the central region, undergoes severe oscillation radially, and may decay to zero beyond a characteristic bounding radius of only a hundred nanometers. These behaviors are distinctively different from the usual photonic angular momentum states in dielectric or metallic materials, which are subject to various diffraction limits. We envision that the extraordinary plasmonic angular momentum states existing in such a minute nanoscale may have broad applications in photonic manipulation \cite{KrishnamoorthyScience2012,ChangPRL2006,KlimovPRA2004}. More thorough studies to this system will constitute our future works.

\section*{Acknowledgements}
We acknowledge the financial support by NSF (ECCS Award No. 1028568) and AFOSR
MURI (Award No. FA9550-12-1-0488).


\begin{thebibliography}{99}

\bibitem{SmithPRL2003} D. R. Smith and D. Schurig, ``Electromagnetic wave propagation in media with indefinite permittivity and permeability tensors,'' Phys. Rev. Lett. {\bf 90}, 077405 (2003).

\bibitem{SmolyaninovPRL2010} I. I. Smolyaninov and E. E. Narimanov, ``Metric Signature Transitions in Optical Metamaterials,'' Phys. Rev. Lett. {\bf 105}, 067402 (2010).

\bibitem{YaoPNAS2011} J. Yao, X. Yang, X. Yin \textit{et al.}, ``Three-dimensional nanometer-scale optical cavities of indefinite medium,'' Proc. Natl. Acad. Sci. {\bf 108}, 11327--11331 (2011).

\bibitem{YangNaturePhoton2012} X. Yang, J. Yao, J. Rho \textit{et al.}, ``Experimental realization of three-dimensional indefinite cavities at the nanoscale with anomalous scaling laws,'' Nature Photon. {\bf 6}, 450--454 (2012).

\bibitem{KrishnamoorthyScience2012} H. N. S. Krishnamoorthy, Z. Jacob, E. Narimanov \textit{et al.}, ``Topological transitions in metamaterials,'' Science {\bf 336}, 205--209 (2012).

\bibitem{FangScience2005} N. Fang, H. Lee, C. Sun, and X. Zhang, ``Sub-diffraction-limited optical imaging with a silver superlens,'' Science {\bf 308}, 534--537 (2005).

\bibitem{JacobOE2006} Z. Jacob, L. V. Alekseyev, and E. Narimanov, ``Optical hyperlens: far-field imaging beyond the diffraction limit,'' Opt. Express {\bf 14}, 8247--8256 (2006).

\bibitem{LiuScience2007} Z. Liu, H. Lee, Y. Xiong \textit{et al.}, ``Far-field optical hyperlens magnifying sub-diffraction-limited objects,'' Science {\bf 315}, 1686 (2007).

\bibitem{LiNatureMater2009} J. Li, L. Fok, X. Yin \textit{et al.}, ``Experimental demonstration of an acoustic magnifying hyperlens,'' Nature Mater. {\bf 11}, 931--934 (2009).

\bibitem{LiOE2009} J. Li, L. Thylen, A. Bratkovsky, ``Optical magnetic plasma in artificial flowers,'' Opt. Express {\bf 17}, 10800--10805 (2009).

\bibitem{DobrzynskiPRB1972} L. Dobrzynski and A. A. Maradudin, ``Electrostatic edge modes in a dielectric wedge,'' Phys. Rev. B {\bf 6}, 3810--3815 (1972).

\bibitem{BoardmanPRB1981} A. D. Boardman, G. C. Aers, and R. Teshima, ``Retarded edge modes of a parabolic wedge,'' Phys. Rev. B {\bf 24}, 5703--5712 (1981).

\bibitem{GarciaMolinaPRB1985} R. Garcia-Molina, A. Gras-Marti, and R. H. Ritchie, ``Excitation of edge modes in the interaction of electron beams with dielectric wedges,'' Phys. Rev. B {\bf 31}, 121--126 (1985).

\bibitem{MorenoPRL2008} E. Moreno, S. G. Rodrigo, S. I. Bozhevolnyi \textit{et al.}, ``Guiding and focusing of electromagnetic fields with wedge plasmon polaritons", Phys. Rev. Lett. {\bf 100}, 023901 (2008).

\bibitem{FerrandoPRE2005} A. Ferrando, ``Discrete-symmetry vortices as angular Bloch modes,'' Phys. Rev. E {\bf 72}, 036612 (2005).

\bibitem{AllenPRB2004} L. Allen, M. W. Beijersbergen, R. J. C. Spreeuw, and J. P. Woerdman, ``Orbital angular momentum of light and the transformation of Laguerre-Gaussian laser modes,'' Phys. Rev. A {\bf 45}, 8185--8189 (1992).

\bibitem{PatersonScience2001} L. Paterson, M. P. MacDonald, J. Arlt \textit{et al.}, ``Controlled rotation of optically trapped microscopic particles,'' Science {\bf 292}, 912--914 (2001).

\bibitem{MarrucciPRL2006} L. Marrucci, C. Manzo, and D. Paparo, ``Optical spin-to-orbital angular momentum conversion in inhomogeneous anisotropic media,'' Phys. Rev. Lett. {\bf 96}, 163905 (2006).

\bibitem{KimNanoLett2010} H. Kim, J. Park, S.-W. Cho \textit{et al.}, ``Synthesis and dynamic switching of surface plasmon vortices with plasmonic vortex lens,'' Nano Lett. {\bf 10}, 529--536 (2010).

\bibitem{ShenOL2012} Z. Shen, Z. J. Hu, G. H. Yuan \textit{et al.}, ``Visualizing orbital angular momentum of plasmonic vortices,'' Opt. Lett. {\bf 37}, 4627--4629 (2012).

\bibitem{YehWaveguide2008} C. Yeh and F. Shimabukuro, \textit{The Essence of Dielectric Waveguides} (Springer, New York,
2008).

\bibitem{HuEPL2012} Q. Hu, D.-H. Xu, R.-W. Peng \textit{et al.}, ``Tune the ``rainbow" trapped in a multilayered waveguide,'' Europhys. Lett. {\bf 99}, 57007 (2012).

\bibitem{LiOE2013} Q. Li and M. Qiu, ``Plasmonic wave propagation in silver nanowires: guiding modes or not?'' Opt. Express {\bf 21}, 8587--8595 (2013).

\bibitem{ChangPRL2006} D. E. Chang, A. S. S{\o}rensen, P. R. Hemmer, and M. D. Lukin, ``Quantum optics with surface plasmons,'' Phys. Rev. Lett. {\bf 97}, 053002 (2006).

\bibitem{KlimovPRA2004} V. V. Klimov and M. Ducloy, ``Spontaneous emission rate of an excited atom placed near a nanofiber,'' Phys. Rev. A {\bf 69}, 013812 (2004).

\bibitem{EconomouPR1969} E. N. Economou, Phys. Rev. {\bf 182}, 539--554 (1969).

\bibitem{JacksonElectrodynamics1998} J. D. Jackson, \textit{Classical Electrodynamics} (John Wiley \& Sons, New York, 1998).

\bibitem{AbramowitzMathFunctions1965} M. Abramowitz and I. A. Stegun, \textit{Handbook of Mathematical Functions with Formulas, Graphs, and Mathematical Tables} (Dover Publications, New York, 1965).

\bibitem{RawlinsPRSLA1999} A. D. Rawlins, ``Diffraction by, or diffusion into, a penetrable wedge,'' Proc. R. Soc. Lond. A {\bf 455}, 2655--2686 (1999).

\bibitem{AshcroftSolidState1976} N. W. Ashcroft and N. D. Mermin, \textit{Solid State Physics} (Thomson Brooks/Cole, Charlotte, 1976).

\bibitem{KikPRB2004} P. G. Kik, S. A. Maier, and H. A. Atwater, ``Image resolution of surface-plasmon-mediated near-field focusing with planar metal films in three dimensions using finite-linewidth dipole sources,'' Phys. Rev. B {\bf 69}, 045418 (2004).

\bibitem{MAOA2010} Y. Ma, X. Li, H. Yu \textit{et al.}, ``Direct measurement of propagation losses in silver nanowires,'' Opt. Lett. {\bf 35}, 1160--1162 (2010).

\bibitem{DavisPRB1976} L. C. Davis, ``Electrostatic edge modes of a dielectric wedge,'' Phys. Rev. B {\bf 14}, 5523--5525 (1976).

\bibitem{McLachlanMathieuFunctions1951} N. W. McLachlan, \textit{Theory and Application of Mathieu Functions} (Oxford University Press, New York, 1951).

\bibitem{BenistyJOSAB2009} H. Benisty, ``Dark modes, slow modes, and coupling in multimode systems,'' J. Opt. Soc. Am. B {\bf 26}, 718--724 (2009).

\end{thebibliography}
\end{document}